\documentclass[preprint,12pt]{elsarticle}

\usepackage{graphicx}% Include figure files
\usepackage{dcolumn}% Align table columns on decimal point
\usepackage{natbib}
\usepackage{placeins}
\usepackage{amsmath}
\usepackage{amssymb}
\usepackage{graphicx}
\usepackage{epstopdf}
\usepackage{xcolor}

\journal{Chaos, Solitons and  Fractals}

\begin{document}

\begin{frontmatter}

\title{Rotating cluster formations emerge in an ensemble of active particles \tnoteref{t1}}
\tnotetext[t1]{This work has been supported by the Spanish State Research Agency (AEI) and the European Regional Development Fund (ERDF, EU) under Project No.~PID2019-105554GB-I00 (MCIN/AEI/10.13039/501100011033)}

\author[1]{Julia Cantis\'{a}n \corref{cor1}}

\cortext[cor1]{Corresponding author}
\ead{julia.cantisan@urjc.es}

\author[1]{Jes\'{u}s M. Seoane}
\author[1,2]{Miguel A.F. Sanju\'{a}n}

\address[1]{Nonlinear Dynamics, Chaos and Complex Systems Group, Departamento de F\'{i}sica, Universidad Rey Juan Carlos \\ Tulip\'{a}n s/n, 28933 M\'{o}stoles, Madrid, Spain}

\address[2]{Department of Applied Informatics, Kaunas University of Technology \\ Studentu 50-415, Kaunas LT-51368, Lithuania}
\begin{abstract}
Rotating clusters or vortices are formations of agents that rotate around a common center. These patterns may be found in very different contexts: from swirling fish to surveillance drones. Here, we propose a minimal model for self-propelled chiral particles with inertia, which shows different types of vortices. We consider an attractive interaction for short distances on top of the repulsive interaction that accounts for volume exclusion. We study cluster formation and we find that the cluster size and clustering coefficient increase with the packing of particles. Finally, we classify three new types of vortices: encapsulated, periodic and chaotic. These clusters may coexist and their proportion depends on the density of the ensemble. The results may be interesting to understand some patterns found in nature and to design agents that automatically arrange themselves in a desired formation while exchanging only relative information.
\end{abstract}

\begin{highlights}

\item Groups of active particles under simple rules form rotating clusters.
\item We find three types of rotating clusters: encapsulated, periodic and chaotic.
\item Our model reflects patterns found in nature and may be used to design artificial agents.

\end{highlights}


\end{frontmatter}


\section{Introduction} \label{Section_1}

Collective motion is an ubiquitous phenomenon that has attracted the attention of mankind for years. The ordered motion of a group of fish or birds that even after a perturbation are capable of reordering to travel together  without the apparent guidance of a leader is a commonly observed example \cite{Katz2011, Couzin2003}. This emergent behavior was modeled by Vicsek and his coworkers \cite{Vicsek1995} using self-propelled particles with an aligning interaction with its neighbors. Many other interesting collective phenomena such as vortices \cite{Delcourt2016}, motility induced phase separation (MIPS) \cite{Cates2014} or synchronization \cite{Pikovsky2021, Chen2017} have also been studied.

The individual constituent particles of systems that exhibit collective motion are active, which means that they are able to take energy from their environment and convert it into directed motion. This implies that these systems are out of equilibrium and must be studied in the context of nonequilibrium physics. These self-propelled particles may have a biological or artificial origin and their size ranges from nanoscopic to macroscopic scale. For biological organisms, self-propulsion is a common feature that improves the exploration of the environment, as it is the case of spermatozoa \cite{Bechinger2016}, bacteria \cite{Ben-Jacob2006} or animals like pine processionary caterpillars or ducks \cite{Delcourt2016}. There is also a growing interest in replicating the collective response of active organisms to construct metamaterials \cite{Klotsa2019} and robots to perform tasks such as drug delivery \cite{Bunea2020}. In the case of man-made active particles, the propulsion is via phoretic transport due to the self-generation of chemical, electrostatic or thermal field gradients. See \cite{Bechinger2016} for a review of artificial active particles. Furthermore, in an engineering context, active particles may refer to mobile agents that explore a region of space for data collection purposes \cite{Freitas2021, Jain2018, Chen2011}. These may be vehicles, satellites, mobile robots or drones. In a physical context, we highlight dusty plasma, which is formed by dust particles in a weakly ionized gas \cite{Morfill2009}. They are common in space, occuring in such diverse environments as interstellar clouds, interplanetary dust, comets, planetary rings and the mesosphere.

Here, we study a type of collective motion: rotating circular formations, also called vortices \cite{Yang2014, Delcourt2016, Kokot2018, Huang2021}. This is widespread in nature and can be found in all of the organisms mentioned in the previous paragraph and in synthetic active matter like curved polymers \cite{Denk2016}. It is also interesting in the case of mobile agents that revisit places periodically to collect data. This is the case of satellite formations, crop monitoring  or land patrolling \cite{Freitas2021}. This pattern is found for chiral active particles (although it can arise in non-chiral particles confined in certain geometries \cite{Huang2021}), which are particles that are not exactly symmetrical relative to their internal propulsion direction and thus their motion is not a straight line, but a circular trajectory. This is the most common case as any small asymmetry in the particles makes the motion chiral.

The phenomenon of collective motion is caused by the interaction of active particles with one another and its environment. Most models consider volume exclusion (to avoid the overlapping and collision of particles) by taking a repulsion potential. A repulsion interaction alone may lead to various collective behaviors like MIPS \cite{Cates2014} and vortices \cite{Kaiser2013}. In other models aligning interactions are considered \cite{Vicsek1995, Martin-Gomez2018,Nilsson2017, Yang2014}. However, we find in nature that often an attractive interaction is present. For interatomic and intermolecular interaction we find attractive forces such as van der Waals forces.  For biological agents, it is common to see that once they are close enough to detect one another, they stay at a certain distance \cite{Franks2016}. And many algorithms used to control groups of robots are based on the behaviour of groups of animals \cite{Son2017}. Therefore, we consider a minimal model that takes into account both a repulsive (volume exclusion) and attractive interaction via the Lennard-Jones potential. To the best of our knowledge, this is the first study of this kind that considers this type of interaction. As we will show, this gives rise to a new plethora of rotating-cluster formation.

Most of the literature has focused on micrometer sized active particles that self-propel in a low Reynolds number liquid (microswimmers), for which inertial effects can be disregarded. Only recently, the case of larger particles or motion in a low-density environment as a gas (microflyers) has been considered \cite{Lowen2020, Caprini2021, Dai2020}. Even if the individual particle is microscopic, when it travels in a swarm, the relevant length scale may be that of the swarm rather than the individual \cite{Klotsa2019}. Thus, inertial effects should also be considered. We include inertial effects in our model by considering underdamped equations of motion. Also, we adopt a simplification: we disregard thermal effects, as in \cite{Pikovsky2021, Yang2014}. This is a reasonable description for larger particles, such as microflyers, for which thermal effects are less intense. As a consequence, we do not include noise in our equations and we deal with deterministic active particles instead of the commonly used active Brownian particles. However, we discuss the qualitative effects that noise would have in the cluster formation.

To sum up, we study the collective motion that arises when chiral active particles interact via a Lennard-Jones potential. We show that they tend to form stable groups called clusters that rotate in different ways. Our model reflects some patterns found in nature \cite{Delcourt2016, Yamamoto2021} and may be interesting for the design of artificial active particles that automatically arrange themselves while exchanging only relative information \cite{Freitas2021, Jain2018, Chen2011}.

The article is organized as follows. In Section \ref{Section_2}, we present our model and the numerical techniques used to analyze it. In Section \ref{Section_3}, the collective behavior of the ensemble is presented. We classify and study in detail the different types of rotatory clusters in Section \ref{Section_4}. Finally, we provide some concluding remarks at the end.

\section{Active chiral particles with inertia} \label{Section_2}

The minimal model we propose implies that the particles follow just some simple rules: chirality, repulsion to avoid overlapping and attraction to maintain a cohesive group. Here, we detail how they are implemented.

\begin{figure}[h]
	\begin{center}
		\includegraphics[width=0.75\textwidth]{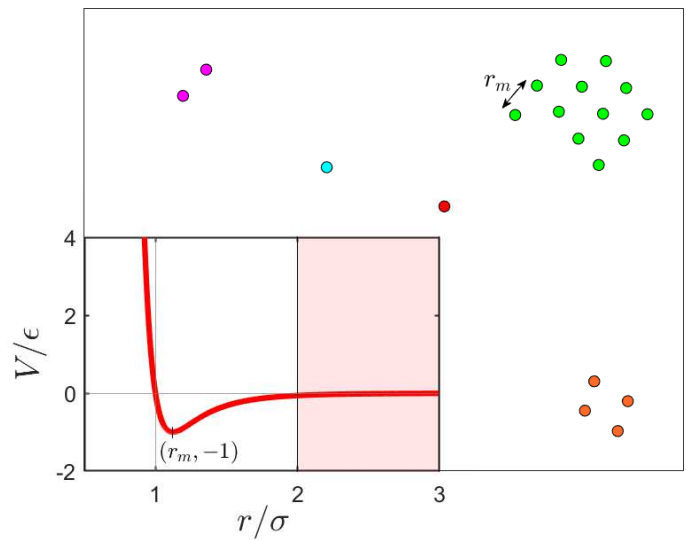}
	\end{center}
	\caption{Steady state for an ensemble of particles at a fixed time. Each group of particles is drawn with a different color as it denotes different clusters. The distance between the closest neighbors in the cluster is $ r_{m} $, which corresponds to the minimum of the potential. In the inset, the Lennard-Jones potential is presented. The shadowed part corresponds to distances beyond the cutoff distance.}
	\label{scheme}
\end{figure}

Our system is formed by an ensemble of active particles that interact with each other and that are trapped in a box with reflecting walls. We consider that particles interact via the pairwise Lennard-Jones potential, which is widely used as a base building block in many interatomic and intermolecular interactions \cite{Wang2019}. The potential depends on the distance $ r $ between the particles in such a way that it is repulsive for short distances to avoid overlapping, while it is attractive when they are further apart, and it vanishes for large distances. The most commonly used expression of the potential is given by the following equation:

\begin{equation}
	V(r)= 4\epsilon  \left[  \left( \frac{\sigma}{r} \right)^{12} -\left( \frac{\sigma}{r} \right)^{6} \right],
\end{equation}
where $ \epsilon $ is the depth of the potential well and it measures the strength of the potential. The distance at which the potential is zero is given by $ \sigma $, providing an idea of how close two particles can get. We fix $ \epsilon=0.1 $ and $ \sigma=0.2 $, so that the behavior is similar to collisions of not-so-hard discs with radius $ \sigma/2=0.1 $ and a cutoff distance at $ 0.4 $ (for which the potential is negligible) to reduce computation times. This cutoff may also be present in biological species for which only their near neighbors are influential and realistic for the design of artificial agents. In the inset of Fig.~\ref{scheme} we represent the potential, where distances beyond the cutoff are shaded. For the parameters chosen both the repulsive and attracting parts are considered. The minimum of the potential is found at $ r_{m}=2^{1/6} \sigma = 0.2245 $, which is the equilibrium distance for two interacting particles.

To define the state of each particle in our $ 2 $D system, it is necessary to ascertain the $ x $ and $ y $ coordinates and the particle orientation $ \phi $. As previously mentioned, we take into account inertial effects and we consider that the particles are chiral. Thus, the system of equations that describes the movement of each particle is:
\begin{equation}
	\begin{aligned}
		 \begin{array}{l}
			m \ddot{x} + \gamma \dot{x}=v \cos{\phi} + F_{LJx} \\
			m \ddot{y} + \gamma \dot{y}=v \sin{\phi} + F_{LJy} \\
			J \ddot{\phi} + \gamma_{R} \dot{\phi}=M,
		\end{array}
	\end{aligned}
	\label{model}
\end{equation}
where $ m $ is the mass and $ J $ is the moment of inertia. Furthermore, $ \gamma $ and $ \gamma_{R} $ are translational and rotational friction coefficients respectively. The first term on the right side for the $ x $ and $ y $ coordinates is the activity term: $ v $ is the self-propulsion velocity and $ \phi $ is the angle of the particle orientation with the $ x-$axis. In the equation for $ \phi $, at the right side we find $ M $, which is the torque responsible for the chirality. The sign reflects if the movement is clockwise or counterclockwise. We take the parameters $ m=\gamma=v=1 $, so that the moment of inertia for our particles (discs with radius 0.1) is $ J=0.005 $ and we fix $ \gamma_{R}=M=0.005 $ too. Finally, the $ F_{LJ} $ terms are the forces that come from the interparticle interaction, in our model the Lennard-Jones potential.

As we are dealing with mesoscopic to macroscopic particles, we disregard thermal effects so that the motion of our ensemble of particles is deterministic. This is in contrast with other studies on active particles that have used the active Brownian particle (ABP) as their minimal model. In fact, our model is a deterministic version of the one in Ref.~\cite{Scholz2018}, in which we replace the noise term by the Lennard-Jones forces. Deterministic models of active particles were previously considered in Refs.~\cite{Pikovsky2021, Yang2014}. However, they did not consider an attracting interaction between the particles. A Lennard-Jones interaction between deterministic active particles has been considered in Ref.~\cite{PRL:2022}.

If we do not consider particle interaction, after some relaxation time, the deterministic particles describe deterministic circular trajectories. The equation for $ \phi $ is decoupled from $x$ and $y$. Thus, it can be directly solved:
	\begin{equation}
		\phi(t)=C_{1} \cdot \frac{J}{\gamma_{R}}  e^{-\frac{\gamma_{R}}{J}t}+ \frac{M}{\gamma_{R}} \cdot t+C_{2},
	\end{equation}
	where $M/\gamma_{R}$ is the spinning frequency $ \omega_{s} $.  For $ t \rightarrow \infty $ the first term corresponding to the transient behavior vanishes. In the steady state, the angle $ \phi $ grows monotonically and the particles are rotating with a constant frequency $ \omega_{s} $. There are two integration constants $ C_{1} $ and $ C_{2} $ that depend on the initial conditions $ \phi_{0} $ and $ \dot{\phi_{0}} $. We have taken $ \phi_{0}=0 $ in all the realizations to make all the particles start with the same phase. However, the clustering behavior we show in this paper is robust for any $ \phi_{0} $ distribution and even for the case that a small noise is added to the equation for the angle and the infinite memory of the initial conditions, present in the deterministic setup, is lost. The system studied is simple and yet it presents a very rich dynamics.

Finally, the spinning radius is:
\begin{equation}
%		\begin{array}{l}
%	\omega_{s}= \frac{M}{\gamma_{R}} \\
	R_{s}=\frac{v \gamma_{R}}{M} \sqrt{\frac{\gamma^{2}}{\gamma^{2}+\left( \frac{M}{\gamma_{R}}\right)^{2} }}.
%	\end{array}
	\label{radius_freq}
\end{equation}

For our parameter values: $ \omega_{s}=1 $ and $ R_{s}=\sqrt{2/3} $. We are interested in a system of particles, each of which follows Eq.~\eqref{model}.

We deal with a low number of particles as we are interested in analyzing in detail how they interact, thus we consider $ N $ between 5 and 35 particles. Our model is formed by $ 6 \times N $  first order ODEs that change each time step as the interaction term is updated with the new positions. In order to integrate it, we employed a Runge-Kutta-Fehlberg algorithm, which is an adaptive step size method, with tolerance $ 10^{-11} $. The algorithm is called for each time step once the interaction forces for the new configuration are calculated. Our particles are confined in a squared geometry of $ L \times L $, with reflective boundaries, so we can define the density of particles as $ \rho=N/L^{2} $. The initial positions and velocities (including $ \dot{\phi_{0}} $) are randomly chosen and we compute at least $ 10^{3} $ runs of the process for each given set of paramaters.

\section{Collective Dynamics} \label{Section_3}

As a result of the interplay between the particles, we find that the system exhibits collective dynamics. For an isolated particle the motion is a well-defined circular trajectory. However, due to the multiple collisions that the particles suffer, the movement becomes chaotic. When we refer to collisions, we do not mean actual collisions as in a hard-spheres model. Rather, we refer to the repulsive interaction via the Lennard-Jones potential that causes that a particle deviates from its initial trajectory.

After multiple collisions and reflections with the walls, the particles adopt a collisionless steady state. In this steady state, particles tend to form clusters, which we define as associations of particles for which its closest neighbors are at a distance equal to $ r_{m} $ and move together as a whole. In Fig.~\ref{scheme}, we can see a snapshot of a run of the process for 20 particles. Each particle is depicted with a different color depending on the cluster it belongs to. We define the transient state as the regime before all the clusters remain invariable for an infinite time if there is no perturbation to the system. Of course, if we considered thermal effects, for certain noise levels the clusters would be constantly created and destroyed and a collisionless steady state would never be reached.

\begin{figure}[h]
	\begin{center}
		\includegraphics[width=0.45\textwidth]{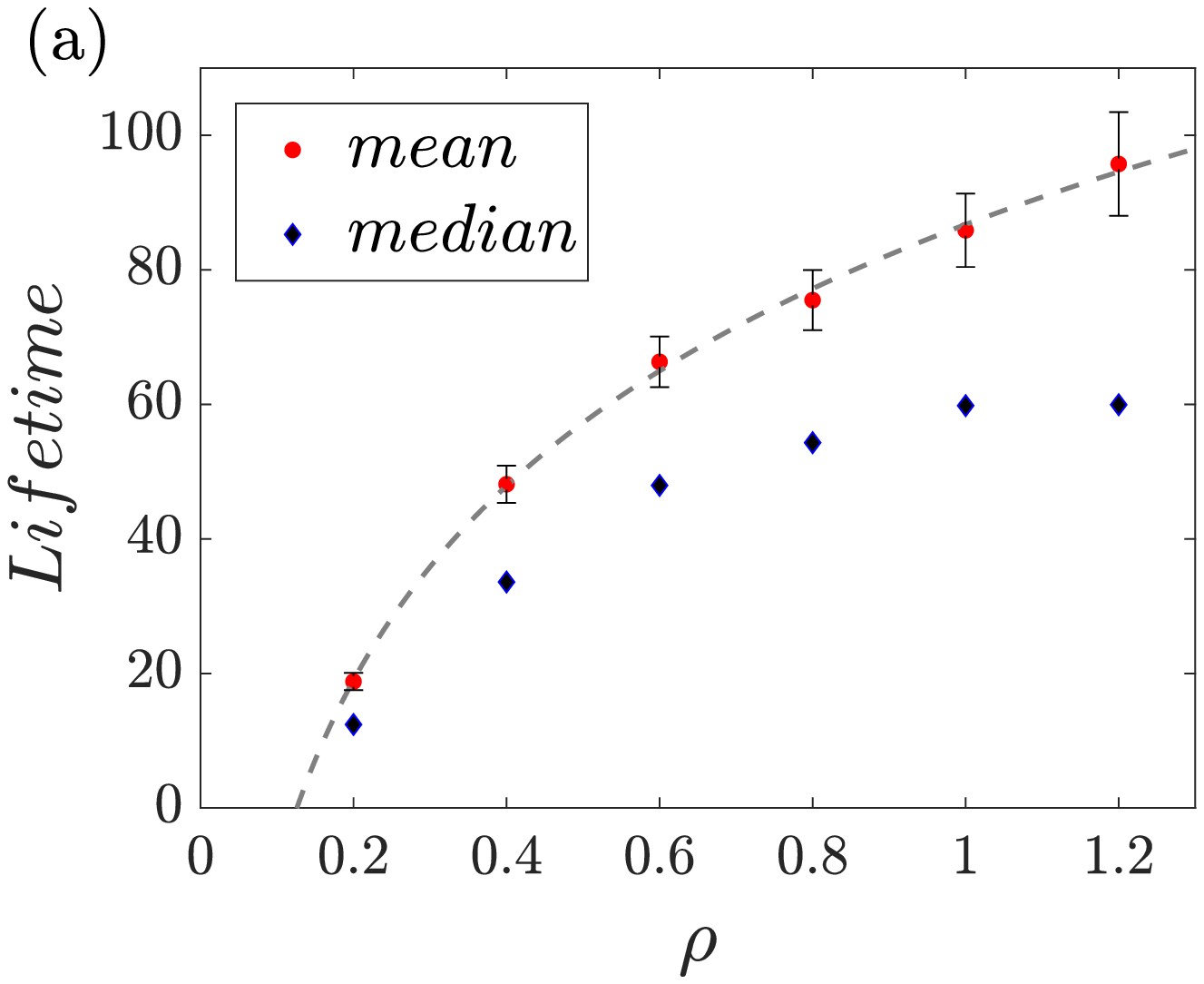}
		\includegraphics[width=0.45\textwidth]{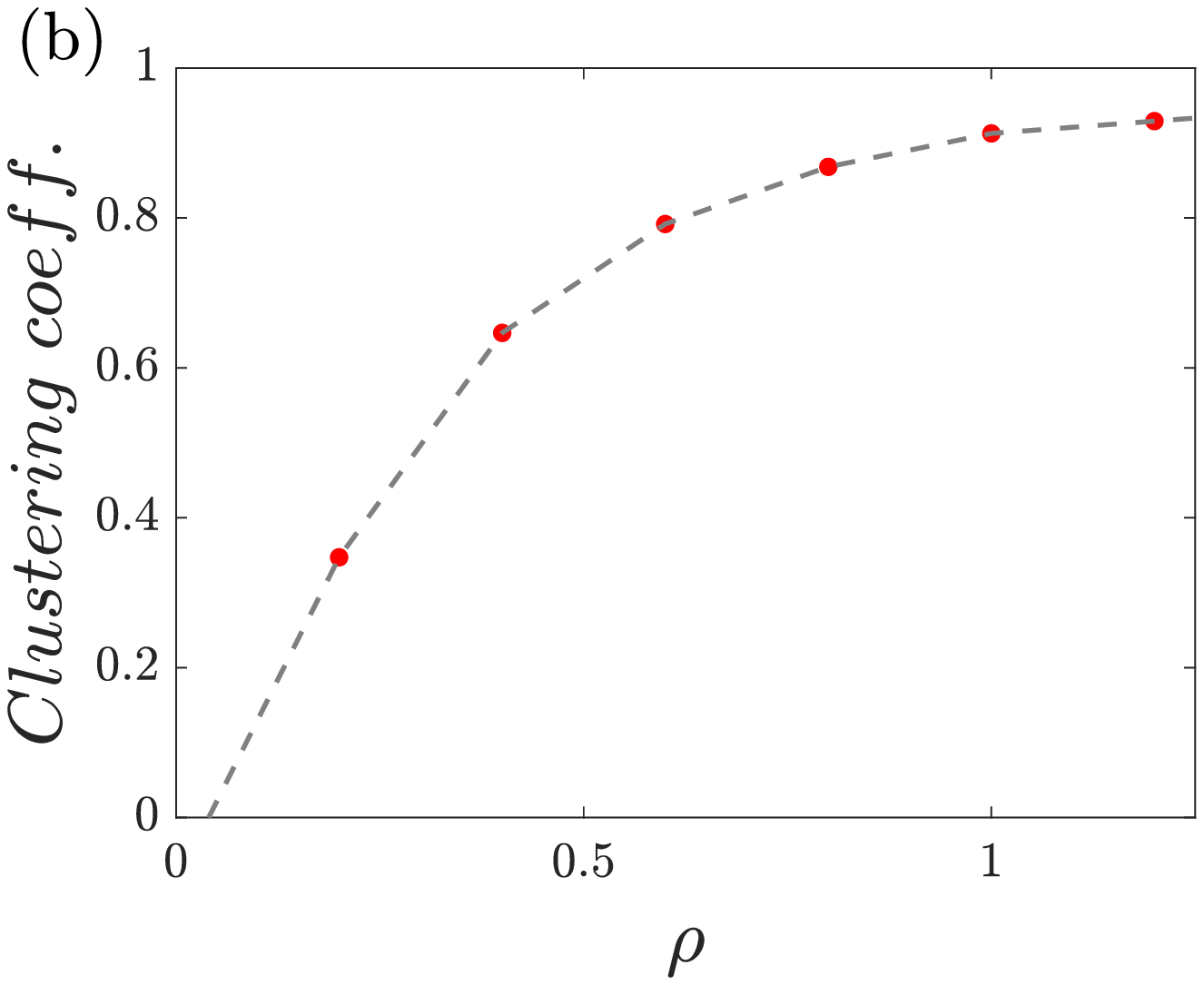}
	\end{center}
	\caption{(a) Transient lifetime dependence with $ \rho $. The mean values are fitted with a power law and are retrieved from $ 10^{3} $ simulation runs. The increasing error bars are caused by an increasing standard deviation as extreme lifetime values tend to occur more frequently for larger values of density. (b) Clustering coefficient dependence with $ \rho $. The values become closer to one (the limit when all particles belong to a cluster) as the density increases.}
	\label{transient_clus_rho}
\end{figure}

In Fig.~\ref{transient_clus_rho}(a), we show the mean values of the lifetime of the transient state for different densities. In this case, we fixed $ L=5 $ and changed the number of particles. The mean values are calculated as a result of multiple runs of the process and the error bars account for the uncertainty due to statistical fluctuations. They depend on the standard deviation ($ STD $) and the number of iterations ($ i $) as $ \varepsilon=2 \cdot STD/ \sqrt{i} $. As expected, an increasing density provokes more collisions and consequently larger transient lifetimes.

The numerically calculated lifetime values ($ \tau $) appear to follow a power law: $ \tau=a \cdot N^b +c$, where $ a=a' \cdot L^{-2b} $ and $ b=0.0225 $,  that is drawn in a gray dashed line in the figure. We do not find supertransients as the transient chaos lifetime does not increase exponentially with the system size. This is in contrast with the previous work by A. Pikovsky \cite{Pikovsky2021}, where more particles mean more collisions and thus exponentially more time to reach a collisionless steady state. However, here the dynamics is modified by the presence of the attracting force that comes from the Lennard-Jones potential. It is the one responsible for the limited growth in the transient lifetime. Also, the median values are represented for comparison. As it can be seen, they increase much more slowly indicating that extreme lifetime values tend to happen for larger number of particles but that the bulk of the trajectories' lifetimes increases slowly.

A useful measure to analyze the degree of collective dynamics is the clustering coefficient, i.e., the fraction of particles that belong to a cluster. As we can see in Fig.~\ref{transient_clus_rho}(b), it grows with density. For low values of $ \rho $ it increases in a more pronounced trend, while for larger values the clustering coefficient comes close to unity, which is the saturation value, and it increases slowly. Studies for other types of ensembles of active particles show that the presence of noise reduces the clustering coefficient. We have to remark that in those cases clusters are metastable. In \cite{Nilsson2017}, they show that a system of active particles with short-range aligning interactions presents a critical noise level at which the transition from a randomized state towards a clustered state occurs. We would expect a similar behavior for our system.

\begin{figure}[h]
	\begin{center}
		\includegraphics[width=0.75\textwidth]{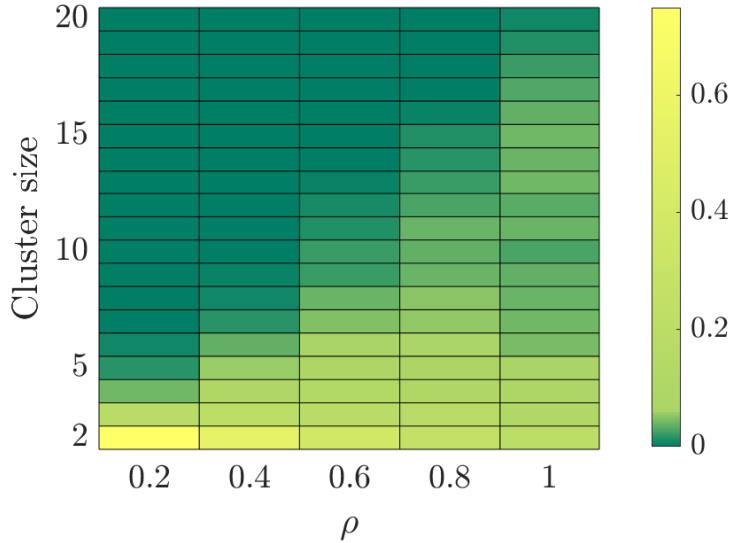}
	\end{center}
	\caption{Cluster size dependence with $ \rho $. For each value of $ \rho $ we present the histogram of cluster sizes (the colorbar denotes the normalized frequency or probability). Clusters become larger when the ensemble is more packed as interactions occur more often.}
	\label{colormap}
\end{figure}

A relevant finding from our model is that not all clusters are the same. In the next section, we present their dynamical differences and we classify them in three groups, but here we talk about the difference in their sizes. When we refer to cluster size, we refer to the number of particles that form a certain cluster. For instance, in Fig.~\ref{scheme} for an ensemble of $ 20 $ particles, after some transient time, we can see clusters of cluster size two, four and twelve. How would this change for other density values? Interactions increase whenever we increase the density, hence one would expect an increase in cluster size. In Fig.~\ref{colormap}, we show how the cluster size depends on $ \rho $. We fix $ N=20 $ and change the box size in order to make the cluster size results comparable. These results are obtained from $ 10^{3} $ simulations, so that for each value of $ \rho $, we plot a histogram of cluster sizes. As it can be seen, there is a transition from low cluster sizes to high cluster sizes. When the packing of particles is small, it is rare to find big clusters as less interactions are probable.

\section{Cluster analysis} \label{Section_4}

Now, we explore the dynamics of the clusters. As the motion of individual particles is chiral, the clusters that are formed are rotational clusters, also called vortices. Unlike what is found in the current bibliography, we find that clusters are not just associations of particles that rotate with the same radius and at the same frequency; their motion is more complex. In fact, we can classify them into three categories attending to their dynamics: encapsulated-rotating clusters (concentric-circular trajectories), periodic-rotating clusters (combination of translation and periodic rotation where both periods are multiples) and chaotic-rotating cluster (combination of translation and chaotic rotation). Different types of clusters may coexist as well as particles that do not belong to any cluster. The formation of a specific type of cluster depends on the velocity and direction of the particles in the moment they come at a distance smaller than the cutoff, thus, cluster formation is a chaotic process.

\subsection{Encapsulated-rotating cluster}

When two or more particles come close to each other, they may form an encapsulated cluster. This is a cluster for which the particles are forced to rotate at a radius different from Eq.~\eqref{radius_freq}. They form concentric circular trajectories that rotate at the same frequency. This is widely observed in nature for animal clustering: see mills in army ants, jack fish or vortices in bacteria \cite{Delcourt2016, Yamamoto2021}.

\begin{figure}[h]
	\begin{center}
		\includegraphics[width=0.75\textwidth]{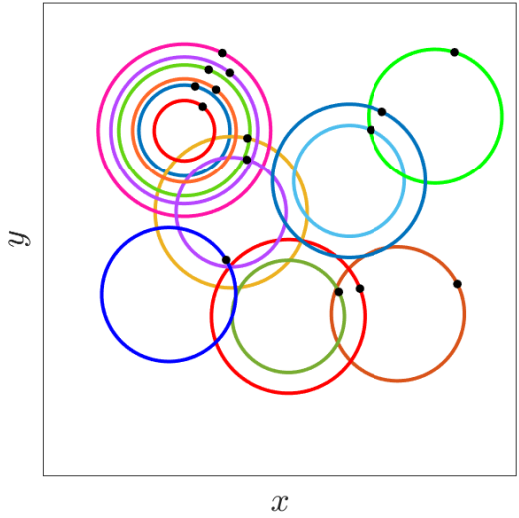}			
	\end{center}
	\caption{Snapshot of the steady state of the system for a realization for which particles are forming encapsulated-rotating clusters. The black points represent the positions at a certain moment of time and the orbits are depicted with different colors. This type of cluster coexists with two isolated particles.}
	\label{encapsulated}
\end{figure}

In Fig.~\ref{encapsulated}, we can see various clusters of this type: one formed by six particles and three formed by two particles. The black points represent the positions at a certain moment of time and the steady-state trajectories are represented with different colors. We call them encapsulated as the external particle in the cluster traps all the others in the inside. These clusters may be formed by any number of particles and depending on the geometry of the clusters, the trapped circular trajectories may have very similar radius (see some of them in the big cluster in Fig.~\ref{encapsulated}, for example), but the particles always remain at a distance $ r_{m} $.

\subsection{Periodic-rotating cluster}

\begin{figure}[h]
	\begin{center}
		\includegraphics[width=0.45\textwidth]{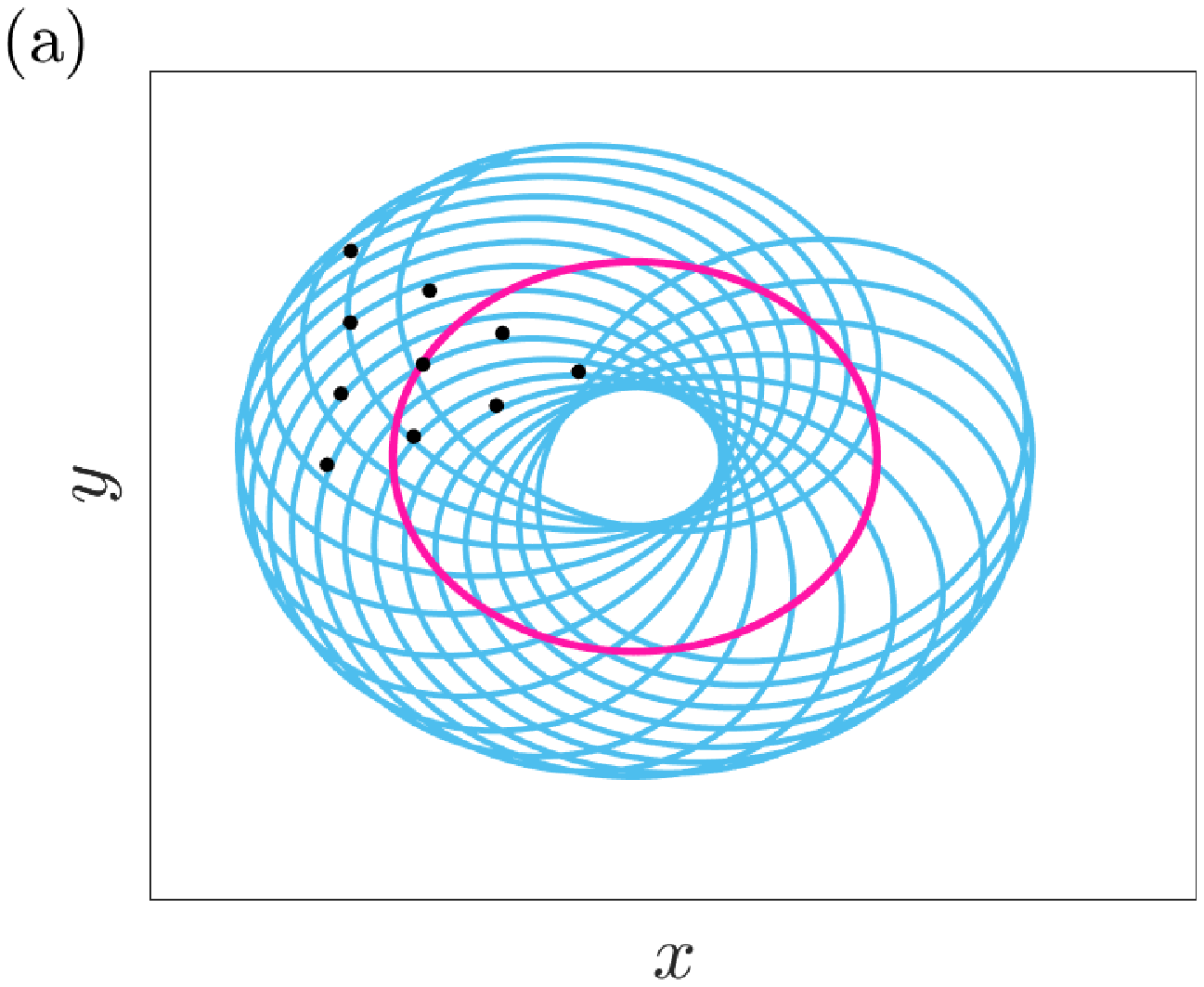}
		\includegraphics[width=0.45\textwidth]{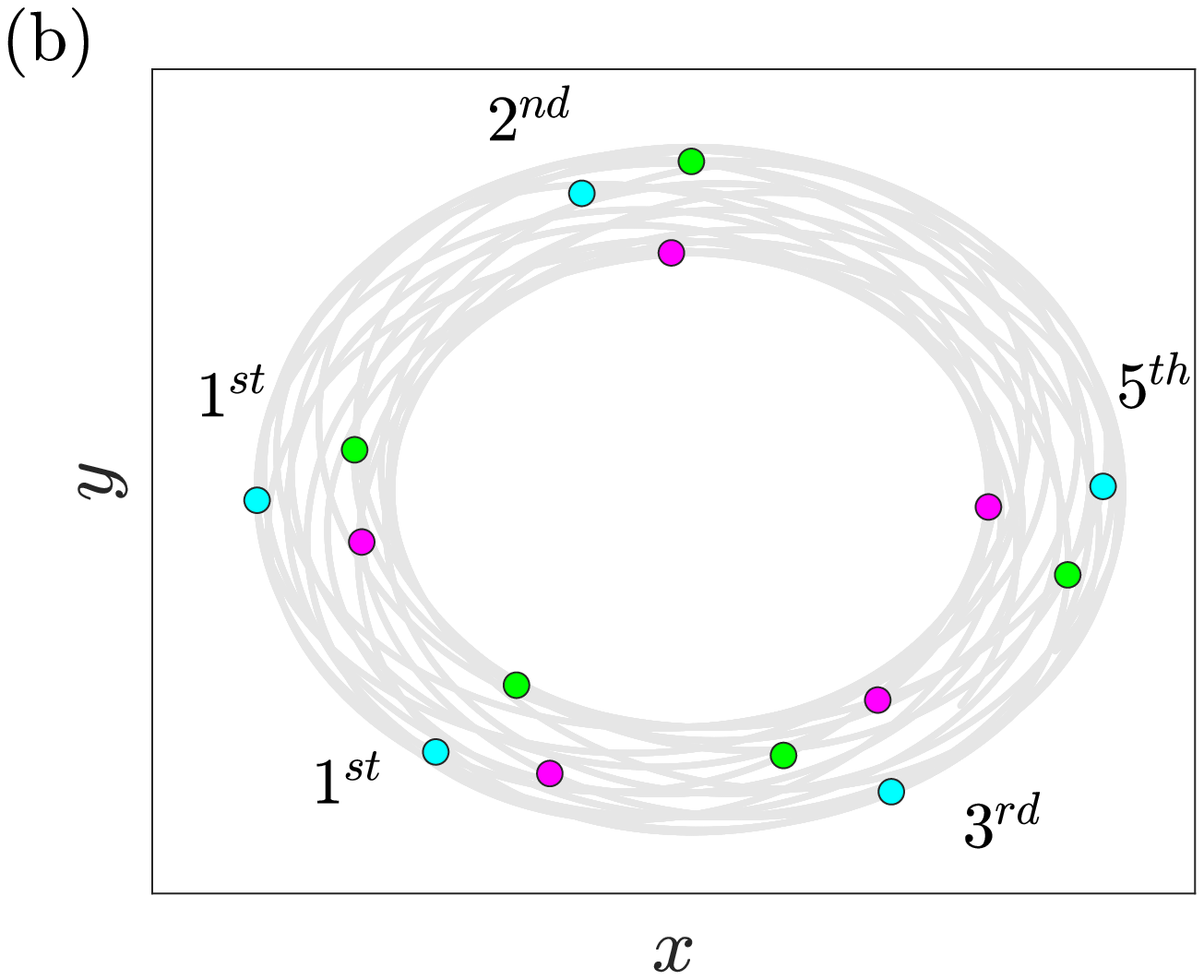}
	\end{center}
	\caption{Representation of the motion of a periodic-rotating cluster. In (a) we show a cluster that rotates slowly compared to the translation. Due to the geometry, the particle in the center describes a circumference (pink trajectory), while the rest describe a more complex motion (blue). In (b) we show a cluster whose period of rotation is similar to that of translation. Particles are represented in different moments of time to observe rotation. The number of the round they correspond to is represented.}
	\label{rotation}
\end{figure}

A more complicated kind of cluster is that of periodic-rotating clusters (PRC). While for encapsulated clusters the outermost particle is always the outermost particle, in PRC this does not hold anymore. In PRC, apart from the circular motion caused by chirality, the cluster itself also rotates. Like the Earth's translation and rotation movements, we find here a combination of both too. For high periods of rotation, that is, when the rotation is slow compared to the translation, we see trajectories like the one in Fig.~\ref{rotation}(a). The cluster is formed by ten particles and the black dots mark the position of the particles at a certain time. In blue we show the trajectory of the particle in the right corner which is performing a circular motion with frequency $ \omega_{s} $ and defined by Eq.~\eqref{radius_freq}  (counterclockwise due to the sign of chirality)  modified by the cluster rotation. We also show the trajectory corresponding to the particle in the center of the triangle (pink trajectory), which describes the translation motion unperturbed by the cluster rotation. This is observed because of the special geometry of the cluster.

When the rotation is faster, what we observe is more similar to Fig.~\ref{rotation}(b), where we can see an example of a cluster formed by three particles. Here, the trajectories are in gray for the three particles and we colored the particles to observe their positions along time. The number of the round they correspond to is marked next to each cluster. The particles are the vertices of a triangle and each particle describes a circular trajectory defined by Eq.~\eqref{radius_freq}. On top of that, the triangle as a whole is also rotating so as in Fig.~\ref{rotation}(a), we have a combination of two rotations.

\begin{figure}[h]
	\begin{center}
		\includegraphics[width=0.75\textwidth]{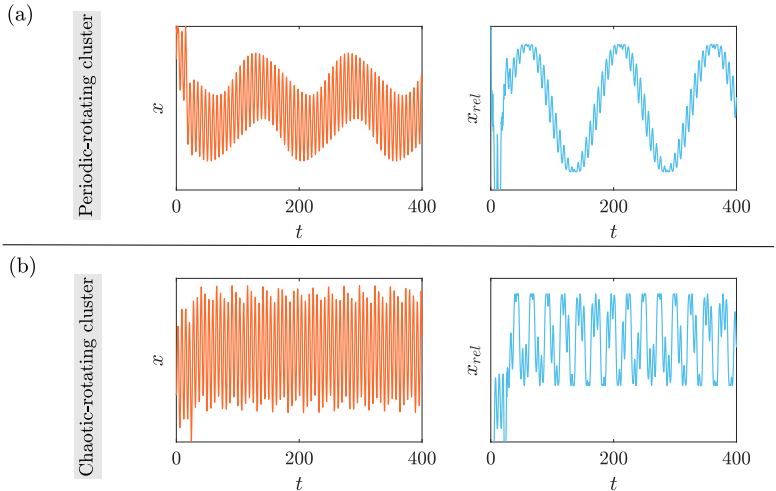}
	\end{center}
	\caption{Time series (left) and relative motion (right) of the $ x- $axis for particles belonging to a periodic-rotating cluster (a) and a chaotic-rotating cluster (b). From the relative motion graphs we have found that even in the periodic case the motion implies a complex forward and backward movement while completing the rotation. }
	\label{plots_type_cluster}
\end{figure}

We say it is a periodic-rotating cluster because there is periodicity in the movement. This fact can be seen in Fig.~\ref{plots_type_cluster}(a) where the time series (left) and the relative motion (right) of two particles from Fig.~\ref{rotation}(a) are shown. From the relative position plot, we deduce that the rotation of the cluster is more complicated than a smooth sinusoidal. There is a forward and backward movement while completing the rotation.

We found numerically that the period of rotation of the cluster, $ T_{cluster} $, is always a multiple of the period of translation ($ T_{s}={\omega_{s}}^{-1} $):
\begin{equation}
	T_{cluster}= \alpha \cdot T_{s},
\end{equation}
where $ \alpha=1, 2, 3... $. For instance, for the cluster in Fig.~\ref{rotation}(b), $ \alpha=7 $, that is, the cluster completes seven rounds before the three particles are in the same position again.

This type of vortex formation may be interesting for engineering applications. This is the case of satellites, vehicles or drones that collect data and may benefit from the combination of the two rotations. This formation allows the system to sample data from different viewpoints, which is necessary for 3D terrain reconstruction and for the cases in which each agent is performing a different task and requires to revisit places periodically.

\subsection{Chaotic-rotating cluster }

Finally, we identified one more type of cluster. Chaotic-rotating clusters (CRC) are clusters for which no periodicity is found in their time series or the relative motion. In Fig.~\ref{plots_type_cluster}(b), we can see the time series (left) and the relative motion of one of the particles from a CRC with respect to another (right). The motion is chaotic and no pattern is found. This type of behavior is suitable for applications that need the agents to develop circular trajectories that are not predictable, such as surveillance, patrolling and local search \cite{Freitas2021, Kafetzis2021}.

\begin{figure}[h]
	\begin{center}
		\includegraphics[width=0.75\textwidth]{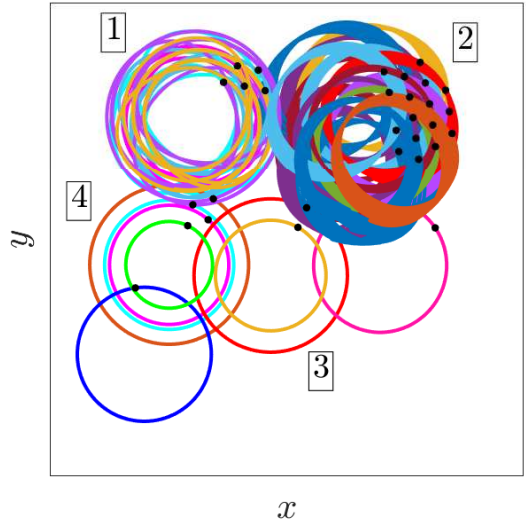}
	\end{center}
	\caption{Snapshot of an ensemble where the black points represent the positions at a certain moment of time and the orbits are depicted with different colors. The three types of clusters coexist: (1) chaotic-rotating, (2) periodic-rotating, (3, 4) encapsulated.}
	\label{type_cluster}
\end{figure}

In Fig.~\ref{type_cluster} we show a run of the process for $ N=30 $ where the three types of the clusters coexist. The clusters are: $ (1)$ chaotic-rotating, $ (2) $ periodic-rotating, $ (3, 4) $ encapsulated. There are also two particles that do not form any cluster. We only show trajectories for the last $ 10\% $ of the computed time. The periodic-rotating cluster is rotating very slowly ($ T_{cluster}>>T_{s} $) and that is why the trajectory is more similar to Fig.~\ref{rotation}(a) than Fig.~\ref{rotation}(b). The chaotic-rotating cluster is not really distinguishable in this phase space representation from a periodic-rotating cluster whose rotation is fast enough. One would have to check the particles time series to observe the difference.

\begin{figure}[h]
	\begin{center}
		\includegraphics[width=0.45\textwidth]{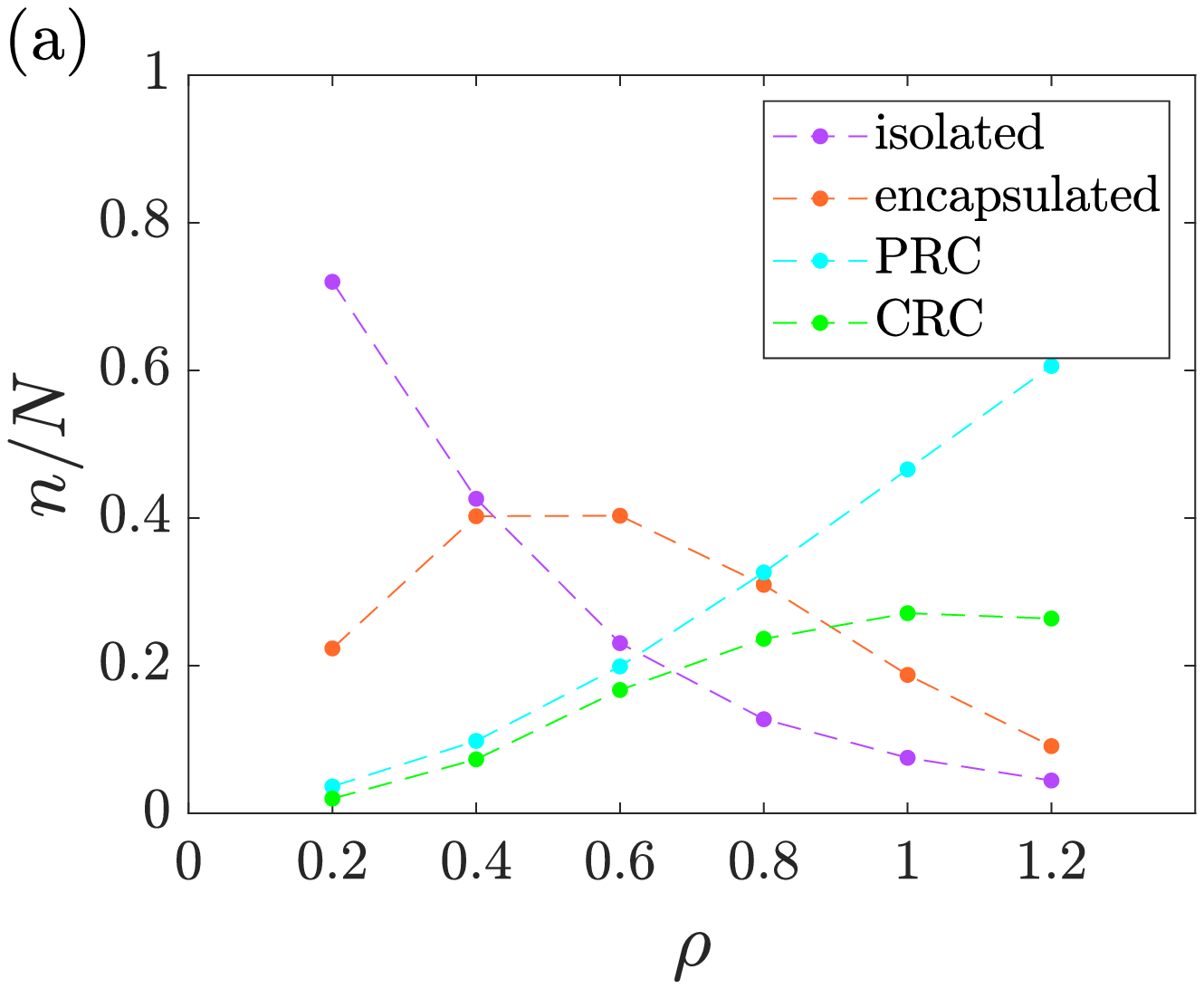}
		\includegraphics[width=0.45\textwidth]{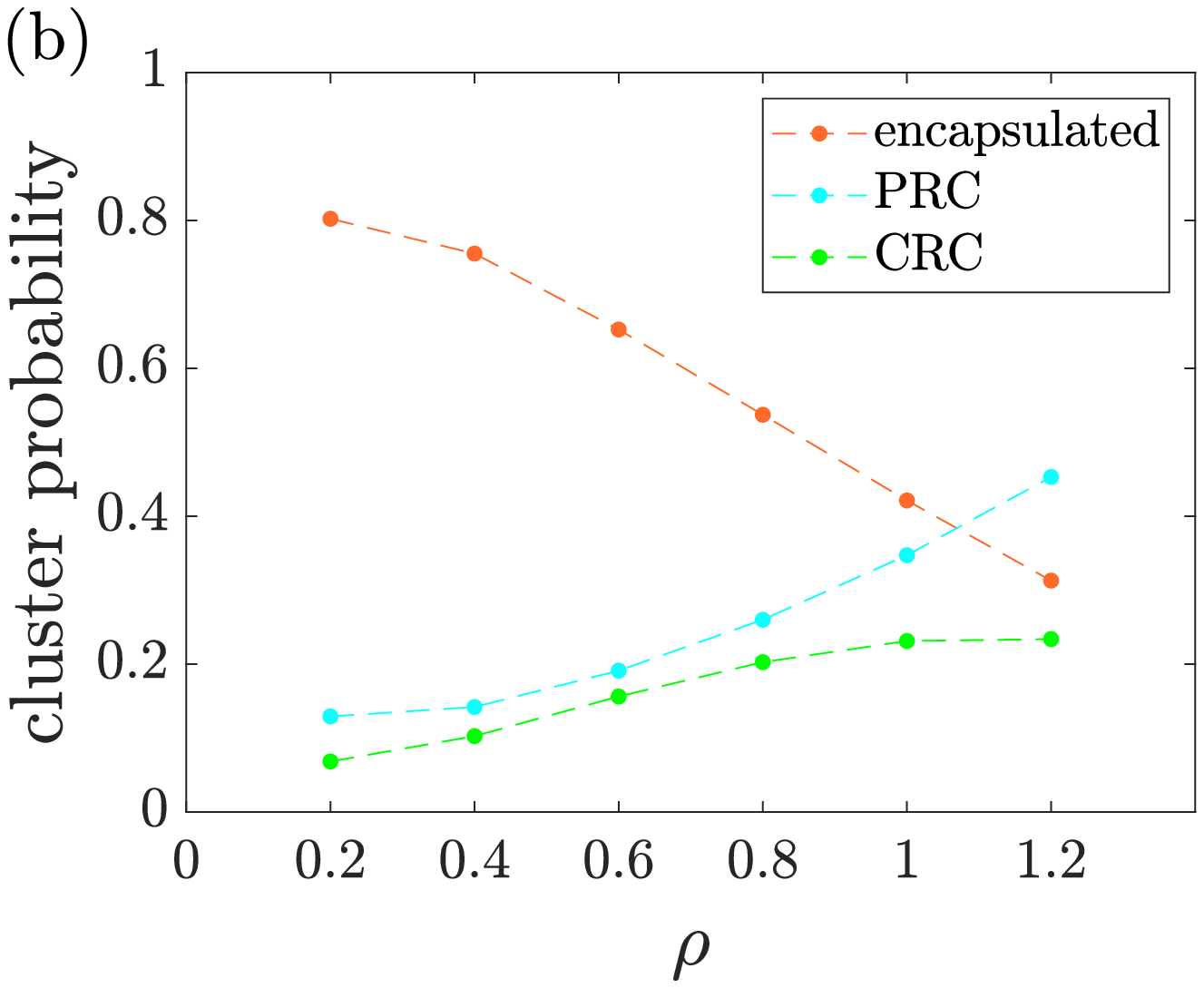}
	\end{center}
	\caption{Evolution of the type of cluster formation with density. In (a) we show the probability that a particle forms each type of cluster, while in (b) we show the probability that a certain type of cluster is formed. These representations differ as the cluster size also changes with $ \rho $. More complex formations (as PRC and CRC) emerge when the density is increased.}
	\label{type_cluster_rho}
\end{figure}

Now that we have classified the different types of clusters, we may ask ourselves how frequently do they appear and how does this frequency depend on density. Do chaotic or encapsulated clusters tend to form more often for high or low densities? We have analyzed that for our three types of clusters. In Fig.~\ref{type_cluster_rho}(a), we show the probability normalized over the total number of particles that a particle forms a certain cluster, averaged over $ 10^{3} $ simulation runs of the process.  The number of particles that do not form a cluster is largely reduced with density, while particles forming PRC or CRC increases. For encapsulated clusters, we find that they reach a maximum over $ \rho=[0.4, 0.6] $ and then decrease.

It is also interesting to compare Fig.~\ref{type_cluster_rho}(a) with \ref{type_cluster_rho}(b). In the latter, we show the cluster probability formation normalized over the total number of clusters in each run, which is different as the cluster size also changes with $ \rho $. For $ \rho=0.2 $ we see that eight out of ten clusters formed are encapsulated. In this representation, this number decreases for $ \rho=0.4$, while in Fig.~\ref{type_cluster_rho}(a) it increases. This is because encapsulated clusters are formed by more particles (bigger cluster size) although the probability of forming an encapsulated cluster is smaller. The probability that an encapsulated cluster is formed, decreases with the density. Interestingly, around $ \rho=1 $ we find that this type of cluster is overcome by periodic-rotating clusters, while the probability for chaotic-rotating clusters is stabilized. In both representations (Fig.~\ref{type_cluster_rho}(a) and \ref{type_cluster_rho}(b)), PRC is the predominant formation for high densities and the appearance of CRC seems to saturate.

Ultimately, what we observe is that for higher values of densities, the complexity of the system increases and more complex clusters are formed.

\section{Conclusions and discussion} \label{Section_5}

Collective motion is the result of the interaction between active particles. A plethora of collective phenomena have been observed in nature and mimicked artificially. Following just a few rules, particles may form vortices or rotating clusters. To that end, we proposed a model that exhibits three novel types of vortices: encapsulated, periodic and chaotic. In our model, we consider active chiral particles that interact via a Lennard-Jones potential (repulsive and attractive) of a larger size than the one appearing  in most of the literature. We do so to cover mesoscopic and macroscopic particles and because even if the individual particle is microscopic, when it travels in a swarm, the relevant length scale may be that of the swarm rather than the individual. Thus, we have included inertial effects and we have disregarded thermal effects, giving rise to deterministic underdamped equations of motion.

We have uncovered a power law that relates the duration of the transient state to the packing of the ensemble of active particles. After this transient state, the system becomes collisionless and forms clusters. We have also observed how the clustering coefficient and the cluster size increase with density.

Furthermore, we have classified the different types of vortices that arise in our model. From less complex to more complex behavior: encapsulated-rotating clusters (concentric circular trajectories), periodic-rotating clusters (combination of translation and periodic rotation, where both periods are multiples) and chaotic-rotating cluster (combination of translation and chaotic rotation). The three types of vortices may coexist and for increasing densities the complexity of the clusters increases.

Finally, our simple model reflects some patterns found in nature for organisms that range from bacteria and spermatozoa to animals and humans. Additionally, it might be interesting for the design of artificial active particles, as robots, satellites or drones that automatically arrange themselves in a desired formation while exchanging only relative information. Some of the tasks that they may perform are tracking, surveillance, sensing and data collection. The results are in like manner promising for the study of active particles in a physical context, as it is the case of dusty plasma.

  \bibliographystyle{elsarticle-num}
  \bibliography{Active_particles_rev}

\end{document}